\documentclass[a4paper,fleqn]{cas-dc}

\usepackage[subrefformat=parens]{subfig}
\usepackage[numbers]{natbib}

\usepackage{graphicx}	
\usepackage{amsmath}	
\usepackage{amssymb}	
\usepackage{booktabs}
\usepackage{times}
\usepackage{microtype}
\usepackage{epsfig}
\usepackage{caption}
\usepackage{float}
\usepackage{placeins}
\usepackage{color, colortbl}
\usepackage{stfloats}
\usepackage{enumitem}
\usepackage{tabularx}
\usepackage{xstring}
\usepackage{multirow}
\usepackage{xspace}
\usepackage{url}
\usepackage{xcolor}

\def\tsc#1{\csdef{#1}{\textsc{\lowercase{#1}}\xspace}}
\tsc{WGM}
\tsc{QE}
\tsc{EP}
\tsc{PMS}
\tsc{BEC}
\tsc{DE}

\begin{document}
\let\WriteBookmarks\relax
\def\floatpagepagefraction{1}
\def\textpagefraction{.001}

\shorttitle{RimSet}

\shortauthors{Jinwei Zhang et~al.}

\title [mode = title]{RimSet: Quantitatively Identifying and Characterizing Chronic Active Multiple Sclerosis Lesion on Quantitative Susceptibility Maps} 
\author[1]{Jinwei Zhang}[]
\cormark[1]
\ead{jwzhang@jhu.edu}

\credit{Conceptualization of this study, Methodology, Software}

\affiliation[1]{organization={Department of Electrical and Computer Engineering, Johns Hopkins University},
    city={Baltimore},
    postcode={MD 21218}, 
    country={USA}}

\author[2]{Thanh D. Nguyen}[]

\affiliation[2]{organization={Department of Radiology, Weill Medical College of Cornell University},
    city={New York},
    postcode={10065}, 
    state={NY},
    country={USA}}

\author[3]{Renjiu Hu}[]

\affiliation[3]{organization={Sibley School of Mechanical and Aerospace Engineering, Cornell University},
    city={Ithaca},
    postcode={NY 14850}, 
    country={USA}}

\author[2]{Susan A. Gauthier}[]
\credit{Data curation, Writing - Original draft preparation}

\author[2]{Yi Wang}[]

\author[4]{Hang Zhang}
\credit{Conceptualization of this study, Methodology, Software}
\cormark[2]
\ead{hz459@cornell.edu}

\affiliation[4]{organization={Department of Electrical and Computer Engineering, Cornell University},
    city={Ithaca},
    postcode={NY 14850}, 
    country={USA}}

\cortext[cor1]{Jinwei Zhang}
\cortext[cor2]{Hang Zhang}

\begin{abstract}
Background and Purpose: Chronic active lesions in multiple sclerosis (MS), termed as rim+ lesions on Quantitative Susceptibility Mapping (QSM), are characterized by a hyperintense rim at the lesion edge and correlate with increased disability in patients. 
While QSM is sensitive to the detection of such lesions, a quantitative characterization of the rim has not been elaborated in literature. 
In this study, we introduce RimSet, which comprises a specially devised set of measurements aimed at quantitatively identifying and characterizing rim+ lesions on QSM \\
Methods: 
The RimSet is composed of two parts.
The first part is the creation of RimSeg, an unsupervised rim segmentation technique, which leverages level-set methodology and incorporates geometric constraints reflecting the lesion appearance on QSM. 
This process serves to segregate the lesion into high-value (rim) and low-value (non-rim) regions.
The second part involves applying first-order radiomic measurements and the Local Binary Pattern as a texture descriptor to the divided subregions of lesions, ultimately forming the RimSet.
We utilized numerical simulations to generate QSM images of lesions, wherein shells and solid spheres represented rim+ and rim- lesions respectively.
To assess the effectiveness of RimSet, we employed an in vivo dataset consisting of 172 MS subjects, comprising a total of 177 rim+ and 3986 rim- lesions.\\
Results: 
In the simulated dataset, we evaluated our segmentation from RimSeg against the ground truth and achieved a Dice score of 78.7\%. 
The major challenges arose from rim+ lesions that presented a partial rim.
In the in vivo data, at the lesion-level, within a five-fold cross-validation framework, our proposed RimSet successfully detected rim+ lesions with a partial area under the receiver operating characteristic curve (pROC AUC) of 0.808. 
This assessment considered clinically relevant false positive rates of less than 0.1. 
The method also secured an area under the precision recall curve (PR AUC) of 0.737.
When compared with other state-of-the-art methods applied to QSM, RimSet demonstrated superior performance in both pROC AUC and PR AUC metrics.
At the subject-level, we evaluated the discrepancy between the predicted rim+ lesion count and the count annotated by human experts. 
In this regard, QSMRim-Net yielded the lowest mean square error of 0.85 and the highest correlation of 0.91 (95\% CI: 0.88, 0.93).

\end{abstract}



\maketitle

\section{Introduction}

Multiple sclerosis (MS) is a central nervous system disorder characterized by inflammation, demyelination, and progressive axonal and neuronal loss \cite{sahraian2007mri}. 
Chronic active lesions, a specific subtype of MS lesions, are identified by a paramagnetic rim indicative of iron-laden inflammatory macrophages and microglia \cite{dal2017slow,absinta2016persistent,kaunzner2019quantitative,gillen2021qsm}. 
Recent research has demonstrated that these chronic active lesions can develop at any stage of the disease and are more common in patients with severe symptoms \cite{harrison2016lesion,absinta2019association,marcille2022disease,luchetti2018progressive}. 
As a result, there is growing interest in utilizing these lesions as biomarkers \cite{rahmanzadeh2022new} for assessing disease burden and monitoring disease progression.

Recent research indicates that certain GRE imaging methods, such as susceptibility-weighted imaging (SWI) \cite{absinta2016persistent} and high-pass filtered phase imaging \cite{dal2017slow}, can characterize chronic active lesions, but these techniques may produce inaccurate quantification of iron susceptibility due to blooming artifacts \cite{marcille2022disease}. 
In this context, quantitative susceptibility mapping (QSM) \cite{wang2015quantitative,wang2017clinical,de2010quantitative} emerges as a promising imaging approach for GRE data, addressing the deconvolution problem between the magnetic field and tissue's magnetic properties, enabling precise quantification and localization of tissue susceptibility. 
Consequently, QSM identifies chronic active lesions with genuine iron deposition, irrespective of imaging parameters \cite{langkammer2013quantitative,stuber2016iron,wisnieff2015quantitative,chen2014quantitative}. 
Moreover, recent investigations have demonstrated that QSM can more effectively illustrate spatial susceptibility patterns (e.g., shell-shaped versus solid lesions) \cite{eskreis2015multiple} in MS lesions compared to phase-based imaging methods. 
The quantitative nature of QSM presents a unique opportunity for automating the process of identification of chronic active MS lesions; This also allows for the standardization of how such lesions, often referred to as 'rim+', are characterized by clinicians.


In this study, we introduce RimSet, a collection of quantitative measurements specifically designed for precise identification and characterization of rim+ lesions using QSM.
By doing this, we aim to reduce the variability in rim characterization among different expert reader groups, which could potentially enhance its clinical translation.
In this study, we initiate our investigation with the use of first-order radiomic measurements as proposed in previous research \cite{kolossvary2018cardiac,kolossvary2017radiomic}. 
For the purpose of quantitatively delineating the contrast properties of rim+ lesions, our approach involves partitioning each lesion into two distinct areas: the high-value (rim) and low-value (non-rim) regions. 
This division is accomplished using an level set segmentation technique \cite{chan2001active} and integrates geometric constraints reflecting the lesion appearance on QSM. 
This method allows us to discern the statistical characteristics that differentiate rim+ and rim- lesions (lesions not identified as rim+) within these defined regions.

Four previous methods have been established for the identification of rim+ lesions: two leverage phase imaging, and two employ QSM. 
RimNet \cite{barquero2020rimnet,zhang2023deda} utilizes a multi-modal VGG Network \cite{simonyan2014very}, APRL \cite{lou2021fully} uses first-order radiomic features paired with a random forest, while QSMRim-Net \cite{zhang2022qsmrim} integrates convolutional and radiomic features with a synthetic minority sampling layer, and DA-TR \cite{zhang2023deda} develops a sophisticated rim parametrization technique for rim+ lesion identification.
We train our developed RimSet using Xgboost\cite{chen2016xgboost} and evaluate their performance in comparison to these established methods when implemented on QSM. 
This evaluation is conducted on both lesion and patient levels, making use of a combination of simulated and real-patient data.

\begin{figure*}[!t]
	\centering
	\includegraphics[width=2.0\columnwidth]{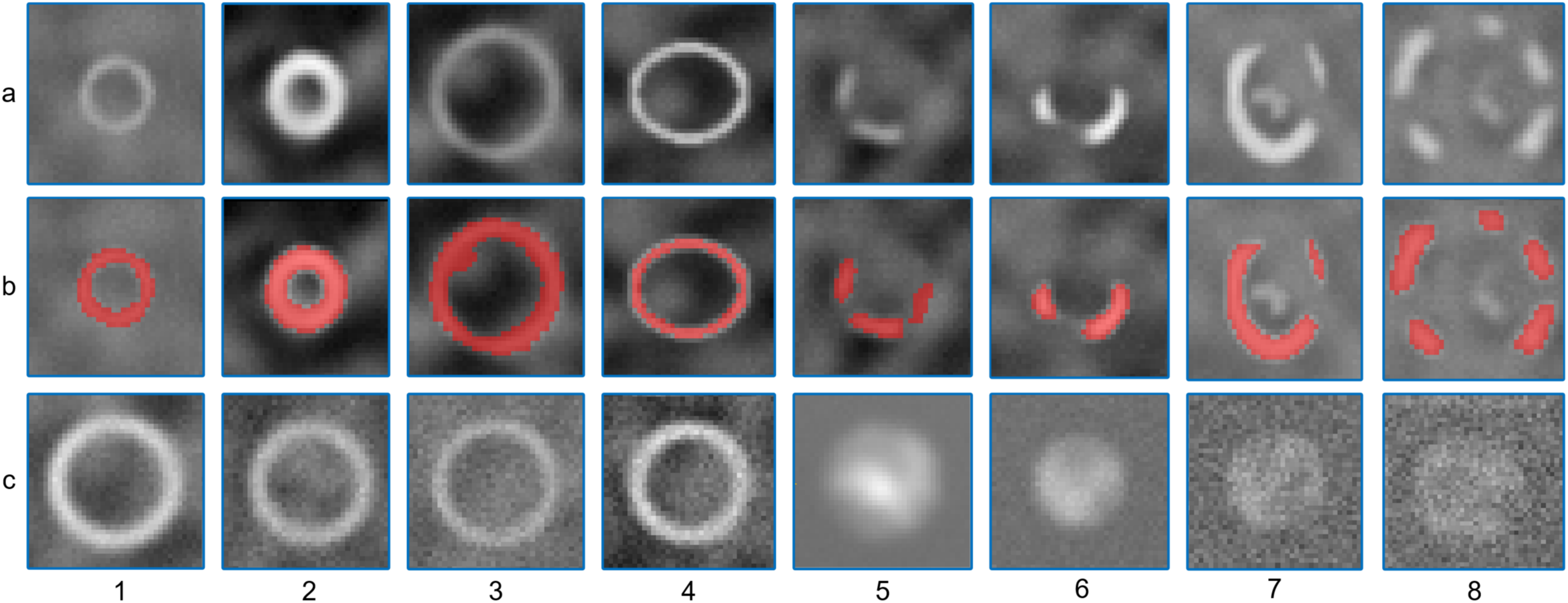}
        \caption{
        Visual representation of the diversity of simulated rim+ and rim- lesions, including their responses to varying noise levels and complex structural characteristics. 
        The first row (a) illustrates variations in the radius, rim thickness, presence of partial rims, oval shape, and vein intersections in the rim+ lesions. 
        The second row (b) displays the corresponding rim segmentation masks derived from Eq.~\ref{eq:levelset-new}, demonstrating its effectiveness in distinguishing rims even in challenging scenarios such as vein intersections. 
        The third row (c) represents rim+ lesions as full shells with increasing noise levels (columns 1-4) and rim- lesions as solid spheres with similar noise augmentation (columns 5-8).
        } 
        \label{fig:simulation}
\end{figure*}

\section{Materials and Methods}

The effectiveness and interpretability of the proposed measurements are validated through a numerical simulation study, which involves various parameter controls for generating lesions. 
These measurements are also assessed using an MS imaging dataset collected at Weill Cornell (refer to Table \ref{tab:demographic}). 
This dataset is composed of data from 172 MS patients who were enrolled in an ongoing prospective database for MS research. 
The database was approved by the local Institutional Review Board and all patients provided written informed consent prior to their inclusion in the database.

\subsection{Datasets}

\subsubsection{Numerical Simulation Data}

\begin{figure}[!t]
	\centering
	\includegraphics[width=1.0\columnwidth]{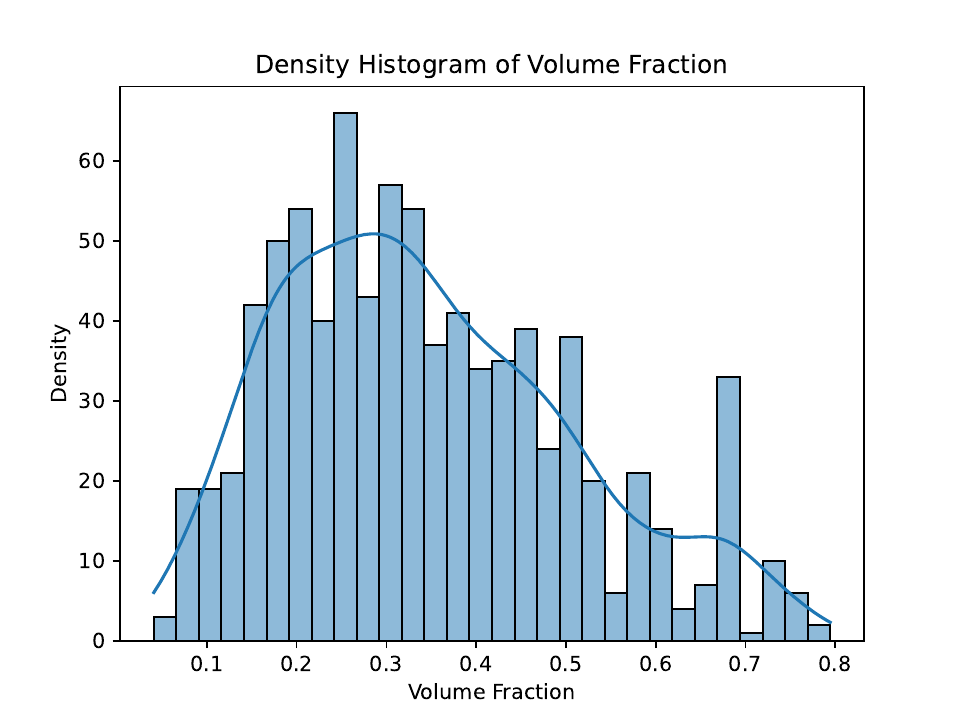}
        \caption{
        A visual illustration of volume fraction of simulated rim+ lesions. 
        The density histogram exhibits a right-skewed, or long-tailed distribution, indicating a diversity of smaller volume fractions and a relatively fewer number of larger fractions.
        } 
        \label{fig:simulation_volume_frac}
\end{figure}

Numerical simulations were employed to produce QSM images of solid spheres and shells, symbolizing rim- and rim+ lesions respectively, utilizing the Python programming language. 
Each lesion was incorporated in an image with dimensions $36 \times 36 \times 12$, and a voxel size of $1 \times 1 \times 3$ mm$^3$. 
We simulated rim+ lesions as shells, varying in characteristics to account for real-world diversity. 
The range for radii spanned from 7mm to 15mm, while thickness varied from 1mm to 3mm. 
The intensity of the high-value region was set within a range of 15 ppb to 45 ppb, and the intensity of the low-value region was constrained between -30 ppb and 0 ppb.
The simulation included rim+ lesions with partial rims, oval-shaped rims, or veins spanning across them to increase the diversity of the data.
Conversely, rim- lesions were simulated as solid spheres, with their radii, higher values, and lower values mirroring those of rim+ lesions. 
To simulate a realistic noise environment, Gaussian noise (mean of 0 and a standard deviation, $\sigma$, ranging from 1 to 7) was applied to both rim+ and rim- lesions. 
Non-lesion background was simulated using simplex noise to ensure smoothness. 
The final image was obtained by seamlessly integrating the lesion within this simulated background.
Following these simulation parameters, we generated a total of 1008 lesions. 
This comprised 840 rim+ lesions, each with ground-truth rim segmentation, which displayed a a right-skewed or long-tailed distribution in terms of rim volume fraction (see Figure \ref{fig:simulation_volume_frac}). 
In addition, the simulation produced 168 rim- lesions.
Please see Figure \ref{fig:simulation} for examples of the simulated data.

\subsubsection{In Vivo MS Data}

\begin{table}[ht]
\centering
\caption{Demographics information for the study cohort.}
\resizebox{1.\columnwidth}{!}{
    \begin{tabular}{ll}
    \hline
    \textbf{Demographics} & \\
    \hline
    Number of Subjects & 172 \\
    \hline
    Gender (count (\%)) & \\
    \quad Female & 124 (72.09\%) \\
    \quad Male & 48 (27.91\%) \\
    \hline
    Disease Subtypes (count (\%)) & \\
    \quad RRMS & 159 (92.44\%) \\
    \quad SPMS & 8 (4.65\%) \\
    \quad CIS & 5 (2.91\%) \\
    \hline
    Disease Durations (mean ± STD) & 10.68 ± 7.37 \\
    \hline
    Age (mean ± STD) & 42.82 ± 10.27 \\
    \hline
    Expanded Disability Status Score (mean ± STD) & 1.38 ± 1.64 \\
    \hline
    Treatment Durations (mean ± STD) & 8.05 ± 5.79 \\
    \hline
    \end{tabular}
}
\label{tab:demographic}
\end{table}

The imaging was performed on a 3T Magnetom Skyra scanner (Siemens Medical Solutions, Malvern, PA, USA). 
The Siemens scanning protocol consisted of the following sequences: 
1) 3D sagittal T1-weighted (T1w) MPRAGE: Repetition Time (TR)/Echo Time (TE)/Inversion Time (TI) = 2300/2.3/900 ms, flip angle (FA) = $8^\circ$, GRAPPA parallel imaging factor (R) = 2, voxel size = $1.0 \times 1.0 \times 1.0$ mm$^3$;
2) 2D axial T2-weighted (T2w) turbo spin echo: TR/TE = 5840/93 ms, FA = $90^\circ$, turbo factor = 18, R = 2, number of signal averages (NSA) = 2, voxel size = $0.5 \times 0.5 \times 3$ mm$^3$; 
3) 3D sagittal fat-saturated T2w fluid attenuated inversion recovery (T2-FLAIR) SPACE: TR/TE/TI = 8500/391/2500 ms, FA = $90^\circ$, turbo factor = 278, R = 4, voxel size = $1.0 \times 1.0 \times 1.0$ mm$^3$. 
For the axial 3D multi-echo GRE sequence for QSM: axial field of view (FOV) = 24 cm, TR/TE$_1$/$\Delta$TE = 48.0/6.3/4.1 ms, number of TEs = 10, FA = $15^\circ$, R = 2, voxel size = $0.75 \times 0.93 \times 3$ mm$^3$, scan time = 4.2 minutes. 
QSM images were reconstructed by the MEDI-0 algorithm \cite{liu2018medi+} from multi-echo GRE data. 
T2-FLAIR images were then preprocessed using the FSL toolbox \cite{smith2004advances}. 
We applied the N4 inhomogeneity correction algorithm to the acquired images and linearly co-registered T2-FLAIR images to the magnitude space of QSM.

Masks for T2-FLAIR lesions were generated for all patients in the dataset. 
These masks were produced by segmenting the T2-FLAIR image with the LST-LPA algorithm \cite{schmidt2012automated}, included in the LST toolbox version 3.0.0 (www.statisticalmodelling.de/lst.html). 
The generated masks were then manually edited and confirmed by the consensus of two expert raters. 
Confluent lesions, which form when pathologically distinct lesions grow close to each other creating a large spatially connected lesion, were identified and subsequently separated and labeled by a human expert. 
Following lesion segmentation and confluent lesion separation, a total of 4,163 individual lesions were identified. These masks were further refined on the QSM image to ensure alignment with the lesions. 
Two human experts manually annotated the rim+ and rim- lesions, scrutinizing each of the 4,163 T2-FLAIR lesions for rim status on the QSM. 
In cases of disagreement, a third human expert provided the final consensus. After the rim lesion annotation, 177 lesions were identified as rim+ lesions and 3,986 lesions as rim- lesions.

\subsection{Segmentation and Measurements}

\subsubsection{Active Contour Model}

Building on prior research \cite{chan2001active}, we have developed an active contour model specifically for rim segmentation. 
The model proposed by Chan and Vese \cite{chan2001active} utilizes piece-wise constant functions to approximate the original image, where the energy function can be minimized by solving a level set evolution equation.
\begin{align}
    & argmin_{c_1,c_2,\phi}F_\epsilon  = \mu\int_{\Omega}{\delta_\epsilon\left(\phi\left(\mathbf{p}\right)\right)}\left|\nabla\phi\left(\mathbf{p}\right)\right|d\mathbf{p} \nonumber\\
    ~ & +v\int_{\Omega}{H_\epsilon\left(\phi\left(\mathbf{p}\right)\right)}d\mathbf{p}  + \int_{\Omega}{\left|\mathbf{\chi}\left(\mathbf{p}\right)-c_1\right|H_\epsilon\left(\mathbf{\chi}\left(\mathbf{p}\right)\right)} \nonumber \\
    ~ & + \int_{\Omega}\left|\mathbf{\chi}\left(\mathbf{p}\right)-c_2\right|\left(1-H_\epsilon\left(\mathbf{\chi}\left(\mathbf{p}\right)\right)\right)d\mathbf{p},
    \label{eq:levelset-ori}
\end{align}
where $\mathbf{\chi}$ represents the input QSM image on a spatial domain $\Omega$. 
The parameters $\mu, v$ are hyperparameters for each fitting or regularization term. 
The function $H_\epsilon\left(z\right)=\frac{1}{2}\left(1+\frac{1}{\pi}arctan\left(\frac{z}{\epsilon}\right)\right)$ is known as the Heaviside function. $\delta_\epsilon\left(t\right)=\frac{dH_\epsilon(t)}{dt}$ is the derivative of the Heaviside function. 
The variable $\mathbf{p} \in \Omega$ denotes the spatial position, $\phi\left(\mathbf{p}\right)$ signifies the value of the level set function at position $\mathbf{p}$, and $\mathbf{\chi}\left(\mathbf{p}\right)$ represents the tissue susceptibility value at position $\mathbf{p}$. 
Finally, $c_1$ and $c_2$ denote the constant susceptibility values above and below the level-set, respectively.

The first term in Eq.~\eqref{eq:levelset-ori} serves as a regularizer for the contour length, where the parameter $\mu$ determines a balance between generating a smoother contour (with a larger $\mu$) and achieving a more accurate image fit (with a smaller $\mu$). 
The second term in Eq.\eqref{eq:levelset-ori} acts as a regularizer for the area enclosed by the contour.
A larger value of the parameter $v$ results in a smaller area within the contour.
In alignment with prior research \cite{chan2001active}, we configured the parameters $\mu=1$ and $v=0.01$ to maintain equilibrium in the segmentation of the foreground and background.

\subsubsection{Regularized Level-Set for Rim Segmentation}

We propose an enhancement to the conventional level-set equation (Eq. \eqref{eq:levelset-ori}) by introducing an additional term that makes use of both the intensity inhomogeneities and the geometric characteristics of rim+ lesions. 
More precisely, we integrate spatial distance information within an exponential function, creating a term reminiscent of temperature, into Eq.\eqref{eq:levelset-ori}.
Let $\mathbf{D}\left(\mathbf{p}\right)$ denote the minimum distance from a voxel located at position $\mathbf{p}$ to the edge of a lesion, and let $D_{max}$ represent the maximum of all $\mathbf{D}\left(\mathbf{p}\right)$ values within the lesion. 
We enhance Eq.~\eqref{eq:levelset-ori} for rim segmentation by replacing $\mathbf{\chi}\left(\mathbf{p}\right)$ with $\mathbf{\chi}\left(\mathbf{p}\right)e^{-w\mathbf{D}\left(\mathbf{p}\right){/D}_{max}}$. 
Here, $w$ is a weight parameter that modifies the significance of the distance in representing the susceptibility value.
This term is designed to favor rim segmentation by attributing greater weight to intensities nearer the boundary. 
This aligns with the geometric structure of rim+ lesions, which are characterized by a hyperintense rim at the lesion's edge. 
Furthermore, by focusing on the lesion's edge, this weighting scheme essentially downplays the influence of inhomogeneities present within the lesion's interior.

With the new term, we can formulate the active contour model specifically designed for rim segmentation as follows:
\begin{align}
    & argmin_{c_1,c_2,\phi}F_\epsilon  = \mu\int_{\Omega}{\delta_\epsilon\left(\phi\left(\mathbf{p}\right)\right)}\left|\nabla\phi\left(\mathbf{p}\right)\right|d\mathbf{p} \nonumber\\
    & +v\int_{\Omega}{H_\epsilon\left(\phi\left(\mathbf{p}\right)\right)}d\mathbf{p} \nonumber\\
    & +\int_{\Omega}{\left|\mathbf{\chi}\left(\mathbf{p}\right)e^{-w\mathbf{D}\left(\mathbf{p}\right){\ /D}_{max}}-c_1\right|H_\epsilon\left(\mathbf{\chi}\left(\mathbf{p}\right)\right)} \nonumber \\
    & +\int_{\Omega}\left|\mathbf{\chi}\left(\mathbf{p}\right)e^{-w\mathbf{D}\left(\mathbf{p}\right){\ /D}_{max}}-c_2\right|\left(1-H_\epsilon\left(\mathbf{\chi}\left(\mathbf{p}\right)\right)\right)d\mathbf{p}.
    \label{eq:levelset-new}
\end{align}

The semi-implicit gradient descent method is utilized to solve Eq.~\eqref{eq:levelset-new}, as detailed in the original paper by Chan and Vese \cite{chan2001active}. 
Solving the eqution yields two constants, $c_1$ and $c_2$, as well as the level set function $\phi$. 
This effectively partitions a lesion into a high-value region with higher susceptibility values and a low-value region with lower susceptibility values. 
We have dubbed this unsupervised rim segmentation technique as "RimSeg". 
Fig. \ref{fig:weight_study} offers visual examples that demonstrate the influence of incorporating spatial distance into the level-set formulation.

\subsubsection{Quantitative Measurements}

Upon the completion of rim segmentation, three distinct masks are generated for each lesion: the full lesion mask, the high-value mask, and the low-value mask. 
Subsequently, we compute a total of 84 measurements that encompass both first-order measurements and a histogram of a texture descriptor.
These measurements encapsulate discriminative information over a lesion, with 13 measurements (including volume, mean, harmonic mean, median, mean absolute deviation (MAD), root mean square (RMS), robust mean square deviation (RMSD), minimum, maximum, the 10th percentile, the 90th percentile, the interquartile range, and the range) describing the average and dispersion of image intensities, 3 measurements (standard deviation, skewness, and kurtosis) characterizing the shape of the image intensity distribution, and 3 measurements (energy, entropy, and uniformity) illustrating the diversity of image intensity.
Furthermore, for each of the three lesion masks, we compute the mean and standard deviation of the distance from each voxel to the lesion edge.
We also calculate the number of spatially separated connected components and volume fractions for the two masks that result from Eq.~\eqref{eq:levelset-new}.
Altogether, these calculations yield a measurement set consisting of 66 unique measurements.

In addition, we incorporated the Local Binary Pattern (LBP) \cite{ojala2002multiresolution,ahonen2004face,ahonen2006face} as an additional texture descriptor to enhance the discriminative power of the RimSet. 
LBP effectively captures the fine-grained intensity variations, proving its usefulness in a variety of image processing applications. 
We computed the LBP feature with a set of 16 circularly symmetric neighbor points and a circle radius of 5. 
To improve rotation invariance and achieve a finer quantization of the angular space, we used the uniform patterns, which is both grayscale and rotation invariant. 
This approach compares the pixel intensity of the central pixel with its surrounding pixels within a circular neighborhood. 
The comparison results are stored in a binary pattern, thus providing rich local texture information. 
This leads to 18 distinct patterns, and for each lesion mask, we calculated the LBP feature, creating a histogram with 18 bins for the texture distribution. 
These histograms were then normalized to form a probability density function, enriching our feature set with local texture details.

Consequently, each lesion on QSM yields a total of 84 measurements, comprising 66 first-order and 18 LBP features. 
Our approach, combining the regularized level-set equation as presented in Eq.~\ref{eq:levelset-new} with these 84 measurements, is denoted as RimSet. 
A detailed enumeration of all first-order measurements is provided in Table 1. 
In this table, $x$ signifies a set of $N_p$ voxels included within the lesion mask. Each voxel is represented by $x_i$, indicating the voxel's intensity. 
The volume of the $i$-th voxel is denoted by $V_i$. 
Furthermore, $p_i$ signifies the normalized first-order histogram with $N_g$ discrete intensity levels, where $N_g$ is the number of non-zero bins. 
The distance from each voxel $x_i$ to the lesion edge is represented by $d_i$.

\begin{table}[h]
    \caption{
        Definitions and formulas for first-order measurements. 
        A vertical line '|' indicates that measurements were calculated based on the masks denoted after the '|'. 
        Absence of a '|' implies that measurements were calculated for all three masks.
    }
    \centering
    \resizebox{1.\columnwidth}{!}{
        \begin{tabular}{ll}
        \hline
        \textbf{Feature} & \textbf{Equation/Definition} \\
        \hline
        Volume & $\sum_{i=1}^{N_p}V_i$ \\
        Mean & $\mu = \frac{1}{N_p}\sum_{i=1}^n x_i$ \\
        Harmonic-Mean & $\frac{N_p}{\sum_{i=1}^n \frac{1}{x_i}}$ \\
        Median & Middle value of intensities \\
        Mean Absolute Deviation (MAD) & $\frac{1}{N_p}\sum_{i=1}^{N_p} |x_i-\mu|$ \\
        Root Mean Square (RMS) & $\sqrt{\frac{1}{N_p}\sum_{i=1}^{N_p} x_i^2}$ \\
        Robust Mean Square Deviation (RMSD) &$\sqrt{\frac{1}{n} \sum_{i=1}^{n} ||x_{i} - \mu||^{2}}$  \\
        Minimum & Smallest intensity value \\
        Maximum & Largest intensity value \\
        10th-Percentile & Value below which 10\% of data fall \\
        90th-Percentile & Value below which 90\% of data fall \\
        Interquartile Range & Difference between 75th and 25th percentiles \\
        Range & Difference between maximum and minimum \\
        Standard Deviation (STD) & $\sqrt{\frac{1}{N_p-1}\sum_{i=1}^{N_p} (x_i-\mu)^2}$ \\
        Skewness & $\frac{1}{N_p}\sum_{i=1}^{N_p} \left(\frac{x_i-\mu}{\sigma}\right)^3$ \\
        Kurtosis & $\frac{1}{N_p}\sum_{i=1}^{N_p} \left(\frac{x_i-\mu}{\sigma}\right)^4 - 3$ \\
        Energy & $\sum_{i=1}^{N_g} x_i^2$ \\
        Entropy & $\sum_{i=1}^{N_g} p_i\log(p_i)$ \\
        Uniformity & $\sum_{i=1}^{N_g} p_i^2$ \\
        Mean-Distance & $ \mu_d = \frac{1}{N_p} \sum d_i $ \\
        STD-Distance & $\sqrt{\frac{1}{N_p-1}\sum_{i=1}^{N_p} (d_i-\mu_d)^2}$ \\
        N-Components | inner \& outer  & Number of spatially-connected components \\
        Volume-Fraction | inner \& outer  & Ratio of the inner to the full mask volumes \\
        \hline
        \end{tabular}
        }
\label{tab:features}
\end{table}

\subsection{Comparator Methods}

Three established methods for identifying rim+ lesions exist: two rely on phase imaging, and one utilizes QSM. 
RimNet \cite{barquero2020rimnet} employs a two-branch VGG Network \cite{simonyan2014very} to extract rim measurements from image patches of phase and T2-FLAIR images. 
APRL \cite{lou2021fully} trains a random forest model on first-order radiomic features extracted from individual lesions of phase images. 
To mitigate the impact of imbalanced data in rim+ lesion identification, APRL also incorporates a synthetic minority oversampling technique. 
QSMRim-Net \cite{zhang2022qsmrim} integrates the strengths of RimNet and APRL by combining convolutional and radiomic features with a synthetic minority sampling network layer. 
The proposed RimSet method is evaluated against these three methods for performance comparison. As APRL and RimNet were initially implemented using phase images, we adapted them to work with QSM for our data.

\subsection{Implementation Details}

For simulation data, we randomly split them into training and testing sets in an 3:1 ratio, which results in 756 lesions for training and 252 lesions for testing.
For in vivo MS data, we applied a stratified five-fold cross-validation procedure to train and validate the performance of the proposed RimSet and the other methods. 
The stratified procedure was performed to balance the number of rim+ lesions in each of the five folds. 
We first grouped subjects into four groups, where the first group contained subjects with no rim + lesion, the second subjects with 1–3 rim + lesions, the third subjects with 4–6 rim + lesions, and the fourth subjects with more than 6 rim + lesions. 
The data was then randomly split into the five folds within each of these groups. 
All experiments were conducted within this stratified five-fold cross validation setting.

RimSet, Xgboost, and other deep neural networks were implemented using Python 3.7 and PyTorch library 1.9.0 \cite{paszke2019pytorch} on a computer equipped with two Nvidia Titan XP GPUs.
The Adam optimization algorithm \cite{kingma2014adam} was employed with an initial learning rate of $1e-3$. 
A multi-step learning rate scheduler reduced the learning rate by half at 50\%, 70\%, and 90\% of the total epochs.
Training utilized a mini-batch size of 32 and was terminated after 50 epochs.
Following previous work \cite{gillen2021qsm}, data augmentation during training included random flipping, affine transformations, and blurring.
For Xgboost, the model was trained using 2000 decision-trees with a maximum depth of 15, and a learning rate set at $5e-3$."

\subsection{Statistical Analysis}

\begin{figure}[!b]
	\centering
	\includegraphics[width=1.0\columnwidth]{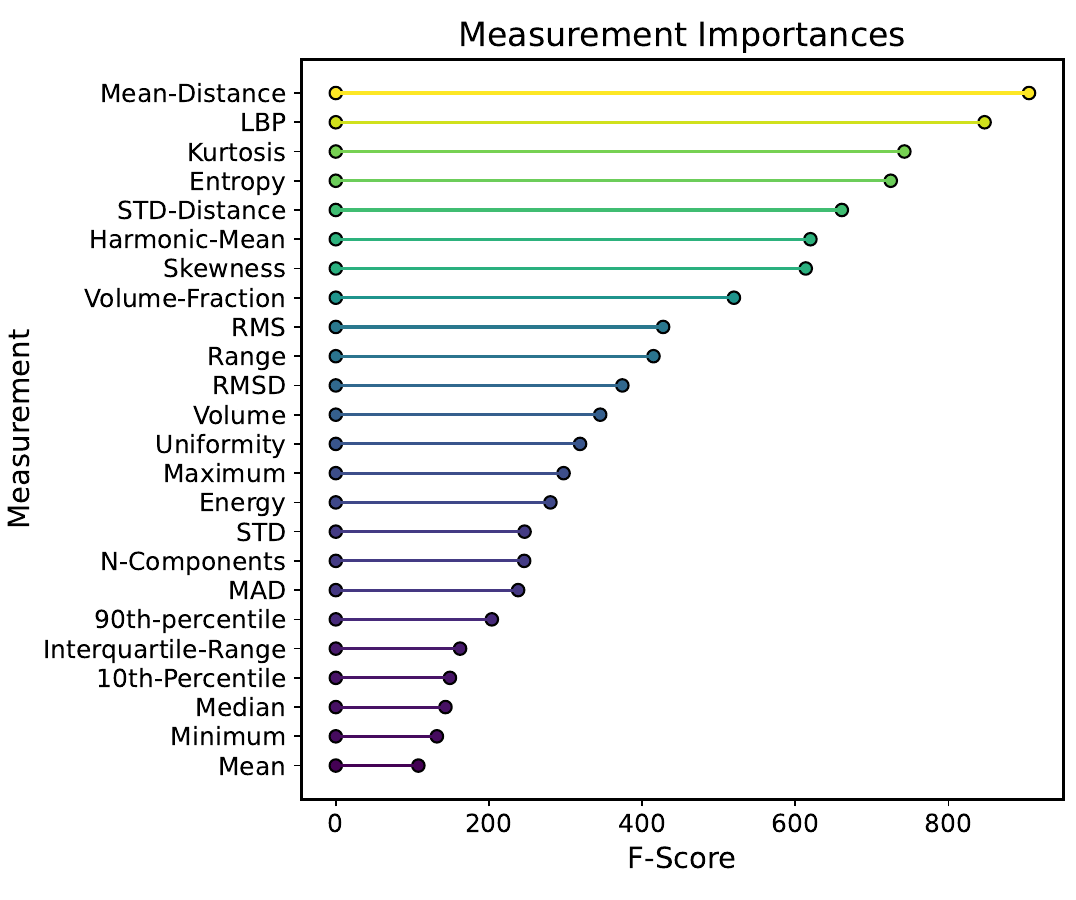}
        \caption{
         A measurement importance plot for the identification of rim+ lesions, derived from Xgboost. 
         A total of 24 measurement, spanning a variety of intensity, texture, shape, and statistical properties, were evaluated. 
         The 'Mean-Distance' was found to be the most influential measurement with a score of 905, followed by 'LBP' and 'Kurtosis' with scores of 847 and 742 respectively. 
         On the other hand, 'Mean' showed the least influence with a score of 107.8. 
         The measurement importance scores offer insights into the discriminative capabilities of these features in the characterization of rim+ lesions.
        } 
        \label{fig:feature_importance}
\end{figure}

\subsubsection{Lesion-wise Analysis}
To assess each method's performance, we interpolated and averaged receiver operating characteristic curves (ROC), partial ROC curves with false positive rates up to 0.1, and precision-recall (PR) curves from different validation folds using piece-wise constant interpolation. 
Given the rarity of rim+ lesions (representing only 4.25\% of total lesions in the in vivo MS dataset), we chose to scrutinize pROC for false positive rates between 0 and 0.1 to obtain more clinically relevant results. 
To differentiate rim+ from rim- lesions, we set a threshold on model probabilities to maximize the F1-score, which is the harmonic mean of precision and sensitivity. 
By prioritizing the F1-score, we achieved a balance between sensitivity and precision as opposed to solely focusing on one. 
Separate thresholds were applied for different folds of the cross-validation, and results were obtained by concatenating the outcomes from the five folds. 
Accuracy, F1-score, sensitivity, specificity, and precision were used to evaluate the performance of each automated method in the lesion-wise analysis.

For the simulated data, we employ the same evaluation metrics as used for the real data. 
Moreover, given the availability of ground-truth rim masks for the simulated data, we also compute the Dice similarity coefficient (Dice Score) between the ground-truth and predicted masks from RimSeg. 
Specifically, we calculate an overall Dice Score for all rim+ lesions, and further compute individual Dice Scores for lesions characterized by either partial or full rims, under increasing levels of Gaussian noise.

\subsubsection{Subject-wise Analysis}
We also assessed performance at the subject-level. 
We used the F1-score criteria for thresholding and compared the number of predicted rim+ lesions and the human expert count number of rim+ lesions for each subject. 
Pearson’s correlation coefficient was used to measure the correlation between the two values. 
Mean Squared Error (MSE) was also used to measure the averaged accuracy for the model predicted count.

\subsubsection{Measurement Analysis}

In order to better understand the characteristics of rim+ lesions, we conducted an analysis of measurement importance. 
XGBoost, rooted in decision tree structures, affords more transparency in understanding how each measurement contributes during the training of the classifier, especially when compared with deep neural networks.
The core concept of XGBoost involves iteratively constructing a set of weak decision trees, each contributing to the overall accurate predictions. 
Decision trees are developed by selecting a feature to divide the data at each stage, with the intention of creating as homogeneous groups as possible. 
Assuming $N$ as the total number of trees in the model, $M_i$ as the number of splits in the $i_{th}$ tree, and $S_{ij}$ as the feature in the $j_{th}$ split of the $i_{th}$ tree, we can obtain the F-score formulation as follows:
\begin{equation}
\text{F-score} (x) = \sum_{i=1}^{N} \sum_{j=1}^{M_i} \mathbb{1}(S_{ij},x),
\label{eq:f-score}
\end{equation}
where $\mathbb{1}()$ is an indicator function assessing the equality of the two inputs. 
Measurements that appear frequently in the decision trees have a higher F-score, while those that are not used at all attain an F-score of zero, hence making F-score an effective indicator of measurement importance.

The ultimate F-score for each measurement listed in Table \ref{tab:features} is derived as follows: for each validation fold, we calculate the F-score for each feature based on the average of the two or three relevant masks (for N-Components and Volume-Fraction, two masks are relevant). 
This score was then averaged across the five folds to obtain the final score. 
For the LBP, we averaged the F-scores from all 18 bins of the histogram for each validation fold, and then, similarly, the final score was obtained by averaging these across the five folds.

\subsection{Ablation Study}

To evaluate the individual and collective impacts of various components in RimSet, we performed an ablation study. 
The study was conducted in a stepwise fashion, beginning with the RimSet configured with only first-order measurements, employing a full mask. 
Next, we modified the RimSet to include just first-order measurements, but utilized all three masks: high, low, and full. 
Following this, we adjusted the RimSet to incorporate only the LBP feature. 
Subsequently, we configured the RimSet to combine both LBP and all first-order measurements. 
In the penultimate setup, we excluded the weighted-distance term from Eq.~\ref{eq:levelset-new}, applying all first-order measurements and the LBP. 
Finally, we integrated the simulated data with the in vivo MS data for training.
This approach was aimed at understanding how the first-order measurements and LBP in the RimSet individually and collectively influence the results.

To further examine the influence of the weight parameter in the level-set equation (Eq.~\ref{eq:levelset-new}), we adjust the weight $w$ and observe its effects on the segmentation of the high-value region of rim+ lesions. 
Specifically, the weight parameter $w$ varies from 0, which corresponds to no weighting on the QSM image, to 0.1, 0.3, 0.5, 1.0, 3.0, 5.0, and finally 10.0. 
We provide visual examples to illustrate how changes in the weight parameter impact the resulting segmentation. 
This investigation helps us understand how the level of influence exerted by the QSM image on the segmentation process can be tuned for optimal results.
\section{Results}

\begin{table*}[!t]

\caption{ Results of the proposed and other methods using a stratified five-fold cross-validation scheme.
The best performing metric is bolded.
}
\vspace{-1ex}
\label{tab:overall}
\begin{center}
\resizebox{2.\columnwidth}{!}{
\begin{tabular}{ lcccccccccc}
\hline
\hline
Method	&Accuracy	&$F_1$ 	&Sensitivity	&Specificity	&Precision	&ROC AUC	&pROC AUC &PR AUC & $\rho$ (95\%CI) & MSE \\
\hline
APRL \cite{lou2021fully}	&0.954	&0.538	&0.627	&0.969	&0.470	&0.940	&0.644	&0.507	&0.68 (0.59,0.75)	&3.16	\\
RimNet \cite{barquero2020rimnet}	&0.970	&0.650	&0.655	&0.984	&0.644	&0.950	&0.737	&0.659	&0.75 (0.67,0.81)	&2.41	\\
QSMRimNet \cite{zhang2022qsmrim}	&\bf{0.976}	&0.711	&0.667	&\bf{0.991}	&\bf{0.761}	&0.939	&0.760	&0.709	&0.89 (0.86,0.92)	&1.00	\\
RimSet  &\bf{0.976} &\bf{0.718} &\bf{0.718 } &0.987 &0.718 &\bf{0.971} &\bf{0.808} &\bf{0.737} &\bf{0.91 (0.88,0.93)}   &\bf{0.85}  \\

\hline
\hline
\end{tabular}
}
\end{center}

\end{table*}

\begin{figure}[!hb]
	\centering
	\includegraphics[width=0.8\columnwidth]{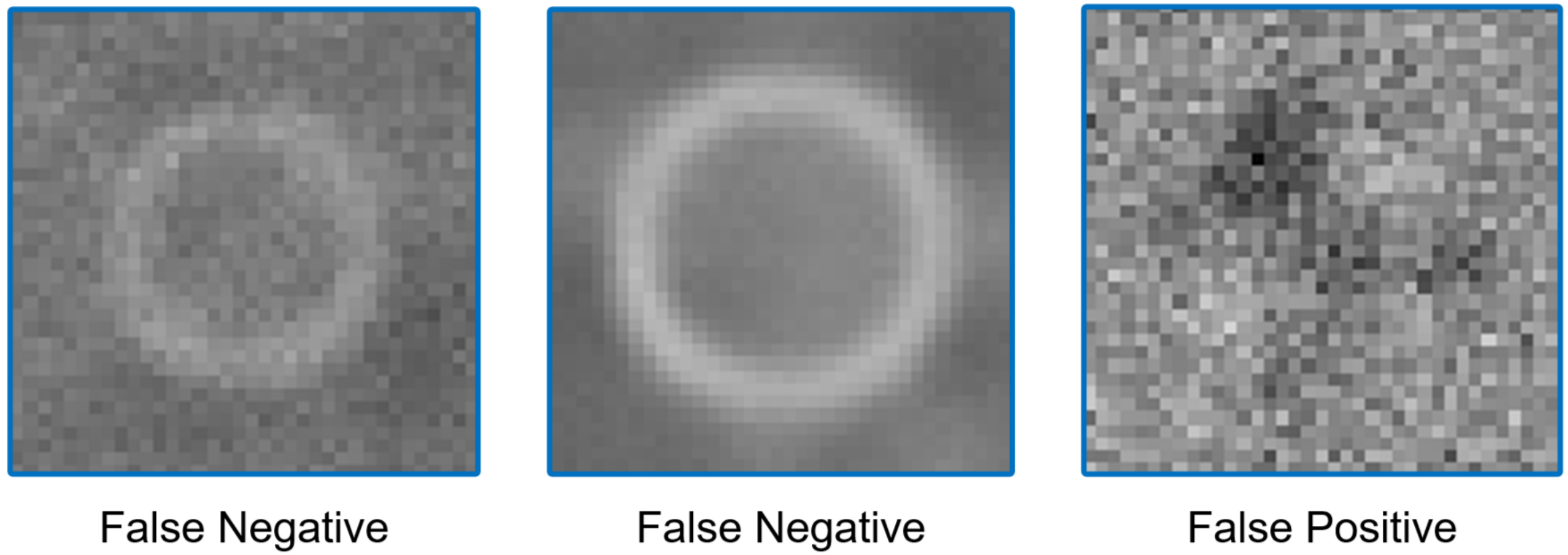}
        \caption{
           Examples of misclassified lesions in the simulated dataset.
        } 
        \label{fig:sim_wrongs}
\end{figure}

\subsection{Results for Simulation Data}

Figure \ref{fig:simulation} presents examples of our simulated data. 
The third row depicts Rim+ lesions portrayed as full shells, with noise incrementally increasing from the first to the fourth column. 
Rim- lesions, simulated as solid spheres with a similar escalation in noise, are displayed in columns five to eight.
The first row illustrates the diversity of rim+ lesions, including variations in radius, rim thickness, the presence of partial rims, oval-shaped lesions, and lesions intersected by veins.
The second row showcases high-value segmentation masks, as generated by Eq.~\ref{eq:levelset-new}. 
The effectiveness of this equation is evident in the accurate rim segmentation, even when a vein bisects the lesion. 
Notably, the vein is not over-segmented, indicating the method's ability to differentiate between structures of similar intensity. 
However, challenges arise when the rim comprises only a minor fraction of the lesion, where background regions with intensities comparable to the rim may be over-segmented.

\begin{figure*}[!t]
	\centering
	\subfloat[ Dice trends against noise]{\includegraphics[width=1.\columnwidth]{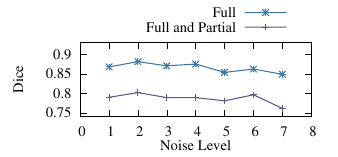}
        \label{fig:dsc_overnoise}} 
	\subfloat[Simulated lesion examples with increasing noise levels]{\includegraphics[width=1.\columnwidth]{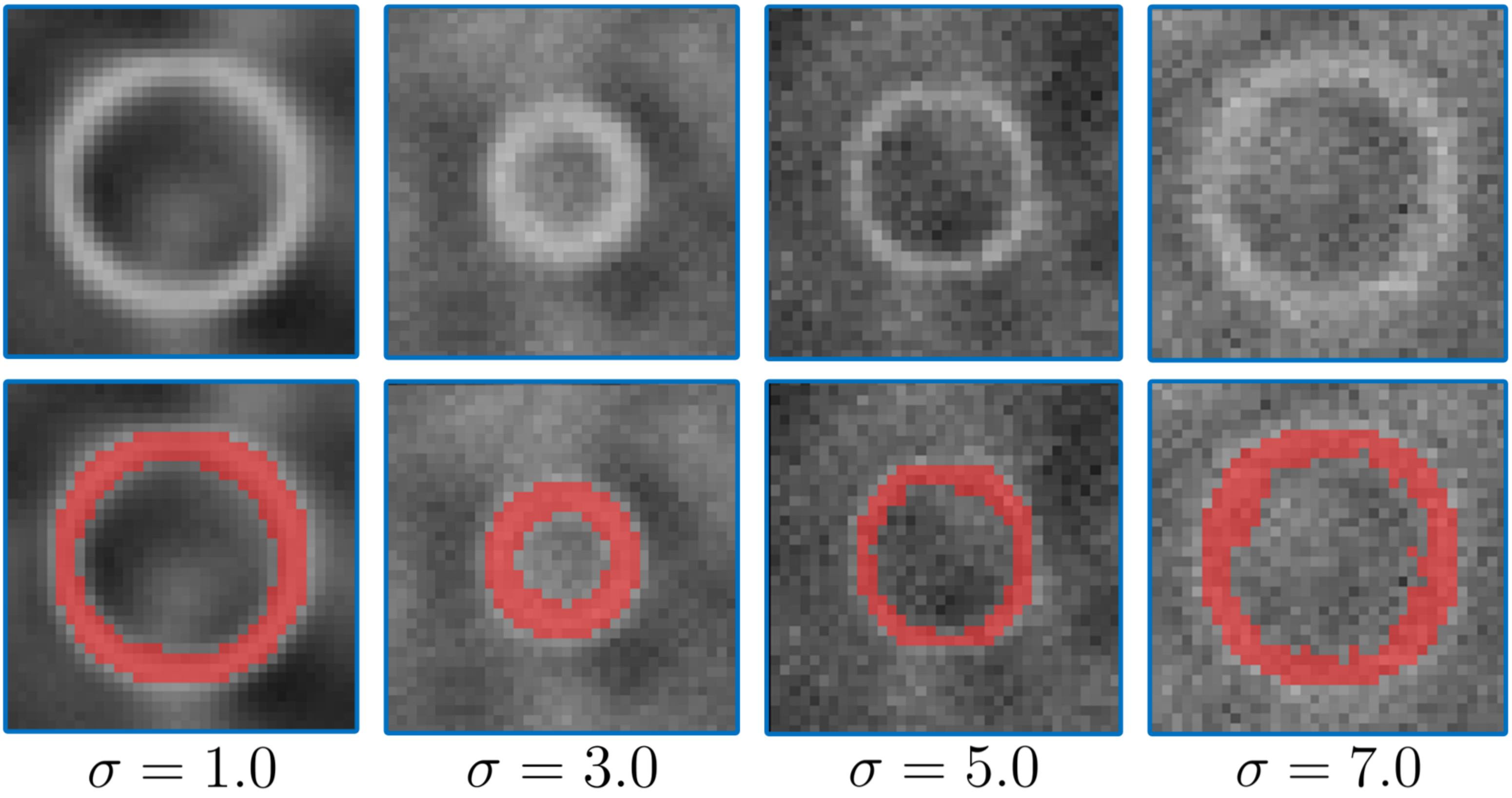}
        \label{fig:noisy_lesions}}
        \caption{
        Performance evaluation of RimSeg on simulated data. 
        (a) presents the robustness of RimSeg against noise by displaying the Dice score trends for rim+ lesions with varying levels of Gaussian noise (sigma). 
        The comparison between full rim and partial rim lesions is also highlighted, showing a performance drop when partial rims are present.
        (b) provides examples of simulated rim+ and rim- lesions with increasing levels of Gaussian noise ($\sigma=1.0, 3.0, 5.0, 7.0$). 
        The effects of increased noise levels on the lesion segmentation by the proposed RimSeg method are illustrated, with over-segmentation observed at a $\sigma=7.0$. 
        } 
\end{figure*}

An average Dice score of 78.7\% was achieved from all 840 simulated rim+ lesions by comparing the ground-truth rim mask with the output masks from the proposed RimSeg method. 
Variation in the Dice score was not significantly impacted by lesion characteristics such as radius, rim thickness, or the presence of an oval shape. 
However, performance was notably affected by the presence of a partial rim. 
As illustrated in Figures \ref{fig:simulation} and \ref{fig:dsc_overnoise}, segmenting rim+ lesions with partial rims was more challenging compared to those with full rims. 
This was reflected in the decline of the average Dice score from 86.5\% for rim+ lesions with full rims to 78.7\% for a combination of full and partial rims. 
Figure \ref{fig:dsc_overnoise} further demonstrates the robustness of RimSeg against noise, with a decrease in the Dice score only occurring when the sigma of the Gaussian noise surpasses 6. 
As Figure \ref{fig:noisy_lesions} reveals, slight over-segmentation was observed when $\sigma$ reached 7.

Using an Xgboost classifier trained on 756 simulated lesions, the model's performance was tested on a separate set of 252 simulated lesions. 
This test resulted in a single false positive and two false negatives, yielding an accuracy of 98.8\% and an F1-score of 99.1\%. 
As depicted in Figure \ref{fig:sim_wrongs}, the false-positive lesion was characterized by extremely high noise, rendering it nearly indistinguishable from the background. 
Conversely, the two false-negative lesions exhibited intensities similar to that of the surrounding background, thus leading to their misclassification.

\subsection{Results for In Vivo MS Data}

\begin{figure*}[!bh]
	\centering
	\subfloat[pROC Curves]{\includegraphics[width=1.\columnwidth]{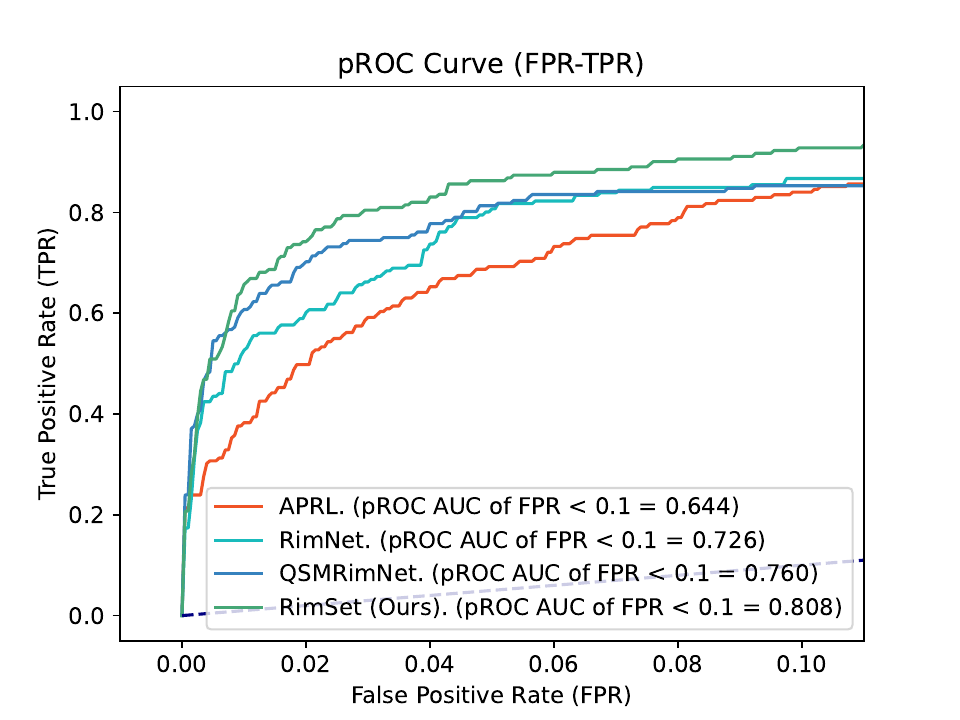}
        \label{fig:pROC}} 
	\subfloat[PR Curves]{\includegraphics[width=1.\columnwidth]{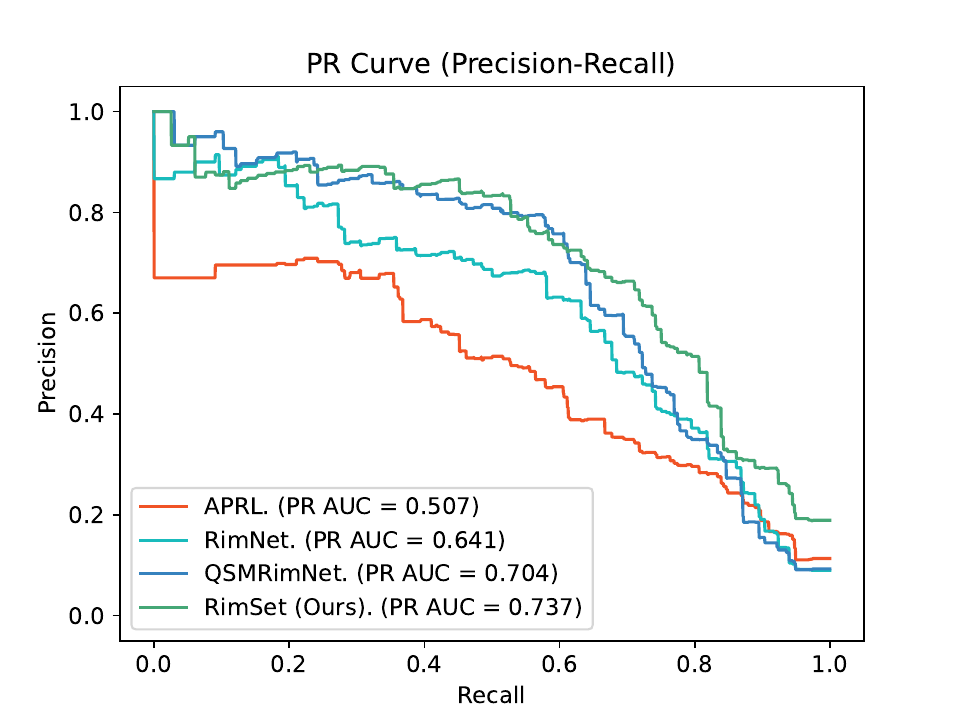}
        \label{fig:PR}}

	\caption{
        The pROC (FPR < 0.1) and PR curves for the proposed and other comparator methods are shown in (a) and (b), where AUC denotes the area under the curve.
	\label{fig:curves}
        }
\end{figure*}

Evaluating the performance metrics of the four methods presented in Table \ref{tab:overall} reveals several noteworthy observations. 
RimSet demonstrated an impressive performance, achieving the highest accuracy of 0.976 and an F1-score of 0.718, reflecting a balanced trade-off between sensitivity and precision. 
On the other hand, QSMRimNet exhibited superior specificity (0.991) and precision (0.761). However, it is important to recognize that while QSMRimNet effectively minimizes false positives, it compromises by missing a significant number of rim+ lesions, leading to an overall lower performance reflected by the curves in Figure \ref{fig:curves}.


Moreover, RimSet achieved the highest ROC AUC of 0.971. 
It also displayed superior performance in terms of pROC (FPR<0.1) AUC of 0.808 and PR AUC of 0.737, indicating that it retains its robust performance in a clinically relevant situation.
Specifically, RimSet resulted in a 3.3\%, 2.2\%, and 3.4\% improvement in ROC AUC, a 25.5\%, 9.6\%, and 6.3\% improvement in pROC AUC, and a 45.3\%, 11.8\%, and 4.0\% improvement in PR AUC compared to APRL, RimNet, and QSMRimNet, respectively. 

In terms of subject-wise analysis, we evaluate the correlation between the predicted count and human expert count using Pearson's correlation coefficient ($\rho$) and MSE. 
RimSet obtained the highest Pearson's correlation coefficient of 0.91 with a 95\% confidence interval of (0.88, 0.93), suggesting a strong positive correlation between predicted count and human expert count. 
Furthermore, RimSet records the lowest MSE of 0.85, indicating that the predicted count of rim+ lesions closely match the human expert count, thereby underscoring its accuracy in individual subject prediction. 

The application of measurement importance analysis using Eq. \ref{eq:f-score} yielded some insightful results. 
As depicted in Figure \ref{fig:feature_importance}, the top 10 measurements in terms of importance are Mean-Distance, LBP, Kurtosis, Entropy, STD-Distance, Harmonic-Mean, Skewness, Volume-Fraction, RMS, and Range.
It's important to emphasize that certain measurements, namely Mean-Distance, STD-Distance, and Volume-Fraction, become particularly meaningful when paired with masks from both high-value and low-value regions.

\subsection{Results for Ablation Study}

The results of the ablation study, as presented in Table \ref{tab:measure_set_eval}, reflect the influences of various components in the RimSet on its performance. 
Six different configurations were examined to assess the individual and combined impacts of the first-order measurements, RimSeg, weighted level-set, LBP, and simulated data.
\begin{table*}[!b]
\caption{The table provides an overview of the ablation study conducted on various components of the RimSet. 
The \checkmark symbol indicates that the corresponding feature is incorporated in the RimSet or that new data is utilized for training. 
Conversely, the $\times$ symbol signifies that the feature is not used. 
The 'Weighted' column reflects the application of the distance-weighted level-set as per Eq.~\ref{eq:levelset-new}.
}
\label{tab:measure_set_eval}
\centering
\begin{tabular}{ccccccccc}
\hline
Model & Full Mask & RimSeg & Weighted & LBP & Simulation & F1 & pROC AUC & PR AUC\\ 
\hline
\# 1 & \checkmark & $\times$ & $\times$ & $\times$ & $\times$ & 0.556 & 0.634 & 0.526\\
\# 2 & \checkmark & \checkmark & \checkmark & $\times$ & $\times$ & 0.606 & 0.699 & 0.611\\
\# 3 & \checkmark & \checkmark & \checkmark & \checkmark & $\times$ & 0.718 & 0.808 & 0.737\\
\# 4 & \checkmark & \checkmark & \checkmark & \checkmark & \checkmark & 0.721 & 0.810 & 0.745\\
\# 5 & $\times$ & $\times$ & $\times$ & \checkmark & $\times$ & 0.635 & 0.729 & 0.593\\
\# 6 & \checkmark & \checkmark & $\times$ & \checkmark & $\times$ & 0.695 & 0.786 & 0.720\\
\hline
\end{tabular}

\end{table*}
Model \# 1, comprising only the first-order measurements with the full lesion mask, achieved an F1 score of 0.556, a pROC AUC of 0.634, and a PR AUC of 0.526. 
In model \# 2, the addition of RimSeg and a weighted level-set to the first-order measurements resulted in improved performance, with the F1 score increasing to 0.606, pROC AUC to 0.699, and PR AUC to 0.611.
A more significant improvement was noted in model \# 3, where the incorporation of LBP in addition to the components of model \# 2 led to an F1 score of 0.718, a pROC AUC of 0.808, and a PR AUC of 0.737. 
A marginal enhancement in performance was observed with the addition of simulated data in model \# 4, which resulted in an F1 score of 0.721, a pROC AUC of 0.810, and a PR AUC of 0.745.
Model \# 5, which contained only LBP, achieved an F1 score of 0.635, a pROC AUC of 0.729, and a PR AUC of 0.593. 
Lastly, model \# 6, excluding the weighted-distance term while including all other components from model \# 3, displayed an F1 score of 0.695, a pROC AUC of 0.786, and a PR AUC of 0.720.
Overall, the results indicate that the combination of first-order measurements, RimSeg, weighted level-set, and LBP yield the best performance, with a slight enhancement provided by the addition of simulated data.

As depicted in Figure \ref{fig:weight_study}, consistent patterns emerge in the rim+ lesions concerning the change of the weight parameter $w$ in Eq.~\ref{eq:levelset-new}. 
As the weight $w$ increases, the high-value region, which targets rim voxels near the boundary, expands in thickness. 
This enlargement peaks when $w=1.0$, after which the region starts to thin out as it becomes increasingly dominated by the distance weight.
\section{Discussion}

In this study, we introduced RimSet, a specially tailored set of measurements aimed at effectively identifying and characterizing rim+ lesions on QSM MRI. 
This represents the first initiative to derive a quantitative measurement set directly from QSM for the discernment and profiling of rim+ lesions, in an effort to minimize variability in rim characterization across different expert reader groups. 
RimSet combines first-order measurements with the LBP texture descriptor, utilizing decision tree-based Xgboost for training and evaluation. 
Impressively, RimSet outperformed leading deep learning-based methods such as QSMRimNet \cite{zhang2022qsmrim} and RimNet \cite{barquero2020rimnet} in terms of F1-score, pROC AUC, and PR AUC.
This emphasizes the potential of RimSet as a robust, precise, and clinically applicable tool for rim+ lesion assessment in QSM MRI.

The RimSet proposed in this work significantly surpasses the performance of APRL \cite{lou2021fully}, a traditional method that also employs first-order radiomic measurements and is trained using a random forest. 
The key distinction between RimSet and APRL lies in our method's ability to integrate observed geometric information from rim+ lesions into the level-set formulation presented in Eq.~\eqref{eq:levelset-new}. 
This enables the division of lesions into high-value and low-value regions.
These subregions within lesions exhibit more discernible characteristics between rim+ and rim- lesions, thus facilitating the identification process. 
Furthermore, by incorporating the texture descriptor LBP, the performance of RimSet exceeds even that of state-of-the-art deep neural network-based methods.
It is worth noting that QSMRimNet combines the benefits of learned convolutional features and radiomic measurements. 
Despite having a tailored DeepSMOTE module to counteract the potential negative impact of the high dataset imbalance, where rim+ has a prevalence of only 4.25\% in our dataset, RimSet still exhibits superior performance.

\begin{figure*}[!t]
	\centering
	\includegraphics[width=2.0\columnwidth]{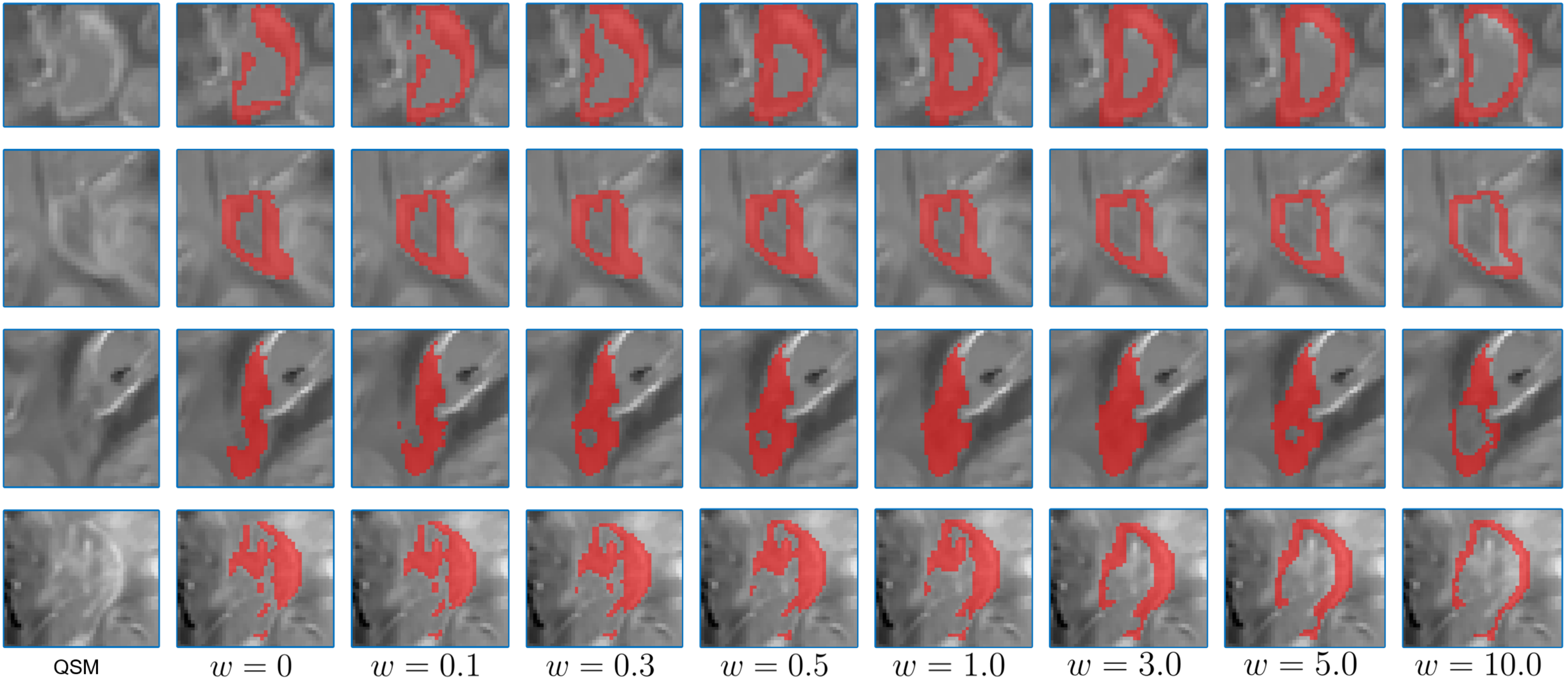}
        \caption{
         Visual representation of the impact of the weight parameter $w$ in the level-set equation (Eq.~\eqref{eq:levelset-new}) on the segmentation of the high-value region in QSM images
        } 
        \label{fig:weight_study}
\end{figure*}

Recent advances in the field of deep learning have significantly enhanced various stages of the lesion analysis pipeline in medical applications. These stages range from image reconstruction, as evidenced by Zhang \textit{et al.} \cite{zhang2023laro,zhangFolded2021}, to critical tasks like image registration \cite{zhang2023spatially,zhang2023dagrid}, lesion segmentation \cite{zhang2019rsanet,zhangGeoloss2021,ZHANG2021102854,ijcai2023p0190}, lesion separation \cite{zhang2021memory}, and lesion identification \cite{zhang2022qsmrim,barquero2020rimnet,zhang2023deda}. 
The fundamental advantage of these deep learning methods is their capacity to automate the feature extraction process through the use of neural network layers. 
However, despite their effectiveness, these methodologies often suffer from a lack of transparency. This opacity in the operational mechanism complicates the interpretation of the models' outputs, especially in the clinical context.
Conversely, our method translates human knowledge and observations of rim+ lesions into a set of scalar measurements. 
This potentially enables broader and more flexible utilization of our technique in clinical environments. 
For instance, as depicted in the top two rows of Figure \ref{fig:weight_study}, when rim+ lesions display a relatively complete rim and significant intensity differences between rim and non-rim regions, RimSeg performs well for different weight parameters. 
Conversely, when a partial rim is present as in the last row of Figure \ref{fig:weight_study}, RimSeg requires fine-tuning of the weight parameter to achieve improved segmentation. 
In light of these conditions, our RimSeg tool offers real-time processing, enabling clinicians to quickly adjust the parameter to achieve the best segmentation map for their particular use case.

The reason we partition lesions into high-value and low-value regions, which respectively represent the rim and non-rim areas, is based on recent findings that susceptibility values at the edge of rim+ lesions are significantly higher than those within the lesion core \cite{kaunzner2019quantitative,marcille2022disease}. 
Furthermore, rim+ lesions often manifest as either a complete or partial hyperintense ring at the lesion's edge on QSM. 
This partitioning strategy enables a more nuanced analysis and a deeper understanding of the disease characteristics. 
Partitioning individual lesions into these high-value and low-value regions essentially presents a binary segmentation challenge, which we address by using active contour models \cite{kass1988snakes,caselles1997geodesic,chan2001active}. 
These models have several advantages over traditional image segmentation methods such as region growing \cite{codella2008left} and minimum cost path \cite{seghers2007minimal}. 
Firstly, active contour models are naturally structured as an energy optimization framework, facilitating the incorporation of prior information, like the spatial intensity distribution of the rim. 
Secondly, active contour models support swift processing, allowing real-time handling of subjects with a high lesion load in less than a second, demonstrating promise for clinical applications.

While the original level-set equation (Eq.\eqref{eq:levelset-ori}) effectively segments rim+ lesions when both the high-value and low-value parts exhibit relatively homogeneous intensities, challenges may arise when accurately segmenting the rim in cases of a hyperintense vein crossing the lesion or more heterogeneous intensity distributions, as shown in the fourth row of Figure \ref{fig:weight_study}. 
Given that the rim is distinctly hyperintense at the edge of the lesion, an approach that mitigates the impact of inhomogeneities caused by hyperintensities in the inner part could enhance the segmentation process.
In our study, drawing upon the geometric structure of the rim, we leverage distance transformation mapping and exponential functions to introduce a temperature-like parameter, referred to as weighted distance, reshaping the susceptibility value for improved rim segmentation. 
This method demonstrated tangible benefits during visual inspection of rim segmentation \ref{fig:weight_study} and also contributed to improved identification performance on comprehensive metrics such as F1-score, pROC AUC, and ROC AUC (see Table \ref{tab:measure_set_eval}). 

Determining whether a voxel in the rim+ lesion from real patient data belongs to the rim or non-rim can prove to be a considerable challenge. 
While visual inspection, exemplified in \ref{fig:weight_study}, assists in comprehending the underlying mechanism, there is a deficiency in a universally accepted gold standard to appraise the accuracy of the segmentation. 
To alleviate this issue, we devised a simulated dataset to aid the process. 
As depicted in Figure \ref{fig:simulation}, the simulated dataset with variations in the radius, rim thickness, presence of partial rims, oval shapes, and vein intersections in the rim+ lesions—provides an effective means to track the accuracy (refer to the Dice trend in Figure \ref{fig:dsc_overnoise}). 
Nevertheless, we identified that the major challenges emerge when a partial rim is present and when the intensity of the rim closely mirrors the surrounding areas. 
Interestingly, our RimSeg proved robust against other variations such as radius, rim thickness, oval shapes, or even vein intersections, with the presence of varying levels of noise. 
It is only when the sigma of Gaussian noise exceeds 6 that we observed a significant performance decline, as well as visual over segmentation. 
These findings align with our observations in real patient data, reinforcing the complexities inherent in rim+ lesion segmentation.

As per the measurement importance analysis depicted in Figure \ref{fig:feature_importance}, the top ten most impactful measurements are Mean-Distance, LBP, Kurtosis, Entropy, STD-Distance, Harmonic-Mean, Skewness, Volume-Fraction, RMS, and Range. 
Additionally, separate analyses were conducted using only the high-value and low-value masks, or the full mask to evaluate their importance. 
Interestingly, in all three scenarios, Mean-Distance consistently ranked as the most crucial feature. 
For the cases exclusively utilizing the high and low-value masks, the maximum intensity emerged among the top ten measurements. 
In contrast, when using the full mask, the 10th-percentile, 90th-percentile, and energy, which represents total contrast weight, became top ten features while others remained consistent. 
This finding suggests that when the lesion is considered as a whole, total contrast weight becomes more significant, indicating that QSM is not only effective at precise depiction, but it also excels at characterizing the lesion from a more global perspective in terms of total susceptibility weight.

Alongside RimSet, our results also benefit the use of QSM. 
QSM offers advantages over phase images by measuring the inherent tissue apparent magnetic susceptibility, thereby enabling the quantification of specific biomarkers, such as iron, that are independent of imaging parameters. 
Previous studies \cite{eskreis2015multiple} have demonstrated that QSM can more accurately depict spatial susceptibility patterns in MS lesions compared to phase-based imaging. 
While shell-like lesions on phase can stem from either solid or shell susceptibility sources, QSM does not present this issue. 
In phase imaging, the presence of a shell-like structure can reflect the magnetic field produced by nearby veins, rather than accurately representing the tissue structure at that point.
The precise localization and quantification of tissue susceptibility offered by QSM empowers us to develop RimSet and RimSeg, allowing for the quantitative identification and characterization of rim+ lesions. 
This aids in reducing the variations in defining rim+ lesions among different imaging centers or expert reader groups.

Despite the promising results, our study has some limitations. 
Both from our simulated and in vivo data, we found that RimSet can struggle to accurately distinguish rim+ lesions with lower susceptibility values, where these lesions often blend into the surrounding white matter, resulting in false negatives. 
Similarly, it has difficulty differentiating rim- lesions with inhomogeneously higher susceptibility values, leading to false positives.
While QSM can precisely quantify tissue susceptibility at a point in space, it's important to note that both iron increase and myelin loss can contribute to the apparent increase of magnetic susceptibility within a lesion on QSM \cite{dimov2022susceptibility}. 
Thus, situations where rim- lesions exhibit extensive demyelination can cause RimSet to produce false positives \cite{rahmanzadeh2022new}, and when rim+ lesions are in the early stages of iron deposition and are between the processes of demyelination and remyelination, they are often classified as false negatives by our RimSet.
Therefore, future studies should consider separating the susceptibility sources of myelin and iron to provide better quantification of rim+ lesions \cite{dimov2022susceptibility,dimov2022magnetic,kim2022subsecond}.

Another limitation of our study pertains to the evaluation of RimSeg's segmentation performance.
In this study, we only utilized the ground-truth segmentation mask from simulated datasets. 
However, as shown in Figure \ref{fig:weight_study}, lesions in real patient data exhibit a more complex appearance with highly inhomogeneous intensity across both the inner and edge of the lesion. 
Thus, it is crucial for human experts to reach a consensus on the rim segmentation on a selection of real patient data and evaluate RimSeg against these masks to ensure the consistency and robustness of the algorithm. 
Therefore, future work should incorporate this additional validation step to enhance the credibility of RimSeg's performance and ensure it meets the rigorous standards of clinical application.

Another limitation of our study is that our RimSet only included first-order measurements and the LBP texture descriptor. 
However, other transformation techniques, such as the Hough Circular Transform (HCT) or Polar Transform (PT), could potentially offer an improved representation of rim+ lesions, characterized by a hyperintense, either partial or full rim. 
For instance, HCT can propagate susceptibility values along the gradient direction, accumulating evidence from the edge of the lesion into the lesion core. 
Moreover, the PT representation is rotation- and scale-equivariant, which may provide more robust quantification of rim+ lesions. 
Therefore, future work should consider integrating these transformation techniques into RimSet, with the aim of enhancing the depiction and characterization of rim+ lesions, ultimately improving the clinical efficacy of the tool.

In conclusion, RimSet and RimSeg, the novel methodologies we have proposed, represent the first set of tailored measurements and segmentation technique designed to effectively identify and characterize rim+ MS lesions on QSM. 
By leveraging a decision tree-based XGBoost model and incorporating active contour models for rim segmentation, RimSet and RimSeg not only outperform other leading deep learning-based methods in rim+ lesion analysis, but also reduce variability in rim characterization among different expert reader groups. 
These methodologies bring a new level of transparency and interpretability to the process, rendering them particularly suitable for clinical applications. 
Hence, RimSet and RimSeg hold good potential to enhance the clinical utility of using the rim+ lesion on QSM as biomarker.

\newpage
\bibliographystyle{plainnat}
\bibliography{cas-refs}

\end{document}